\documentclass[12pt]{article}
 
\advance\voffset by -1.5cm
\advance\hoffset by -2.1cm
\usepackage{epsf}

\def\drawbox#1#2{\hrule height#2pt
        \hbox{\vrule width#2pt height#1pt \kern#1pt
              \vrule width#2pt}
              \hrule height#2pt}

\def\Asym#1#2{\vcenter{\vbox{\drawbox{#1}{#2}
              \kern-#2pt       
              \drawbox{#1}{#2}}}}

\def\href#1#2{#2}
\newcommand{\rot}[3]{\left[{#1}\atop{#2}\right]_{#3}}
\newcommand{\be}{\begin{equation}}
\newcommand{\ee}{\end{equation}}
\newcommand{\bea}{\begin{eqnarray}}
\newcommand{\eea}{\end{eqnarray}}
\newcommand{\ena}{\end{eqnarray}}

\newcommand{\ba}{\begin{array}}
\newcommand{\ea}{\end{array}}
\newcommand{\beann}{\begin{eqnarray*}}
\newcommand{\eeann}{\end{eqnarray*}}
\newcommand{\del}{\partial}
\newcommand{\pa}{\partial}
\newcommand{\e}{\epsilon}
\newcommand{\Z}{{\mathbf Z}}

\newcommand{\Tr}{\mbox{Tr}}

\begin{document}

\thispagestyle{empty}
\def\thefootnote{\fnsymbol{footnote}}
\begin{flushright}
  hep-th/9908075\\
  CALT-68-2235\\
  MIT-CTP-2890\\
  TAUP-2587-99\\
  NSF-ITP-99-093
  \end{flushright}
\vskip 0.3cm

\begin{center}\LARGE
{\bf Branes and Supersymmetry Breaking in 3D Gauge Theories}
\end{center}
\vskip 0.5cm

\begin{center}\large
  Oren Bergman$^{a}$,
  Amihay Hanany$^{b}$,
  Andreas Karch$^{b}$,
  Barak Kol$^{c}$
\end{center}
\vskip 0.3cm

\begin{center}\it$^a$
California Institute of Technology\\
Pasadena, CA 91125, USA\\
e-mail: {\tt bergman@theory.caltech.edu}
\end{center}

\begin{center}\it$^b$
Center for Theoretical Physics\\
Laboratory for Nuclear Science\\
Massachusetts Institute of Technology\\
Cambridge, MA 02139, USA\\
e-mail: {\tt hanany,karch@ctp.mit.edu}
\end{center}

\begin{center}\it$^c$
School of Physics and Astronomy\\
Beverly and Raymond Sackler Faculty of Exact Sciences\\
Tel Aviv University, Ramat Aviv 69978, Israel\\
e-mail: {\tt barak@beauty.tau.ac.il}
\end{center}

\vskip 0.7cm

\begin{center}
August 1999
\end{center}

\vskip 0.7cm

\begin{abstract}

It is shown that supersymmetry is spontaneously broken
in certain three-dimensional supersymmetric gauge theories, 
by using the s-rule in their string theory realization as brane 
configurations. 
In particular, supersymmetry is broken in
$N=3$ supersymmetric Yang-Mills-Chern-Simons 
theory with gauge group $SU(n)$ and CS coefficient $k$, 
as well as in its $N=2$ and $N=1$ deformations, when $n>|k|$.
In addition, supersymmetry is broken in the 
$N=1$ mass deformation of $N=2$ supersymmetric Yang-Mills theory
with gauge group $SU(n)$ and one matter multiplet when $n>1$.
In the latter case the breaking is induced by an instanton-generated 
repulsive potential.

\end{abstract}

\vfill
\setcounter{footnote}{0}
\def\thefootnote{\arabic{footnote}}
\newpage

\section{Introduction}

Much has been learned about gauge theories in various dimensions
and with various amounts of supersymmetry from the study of brane
configurations in Type II string theories.
Examples which stand out are mirror symmetry in three-dimensional
$N=4$ supersymmetric Yang-Mills (YM) theory \cite{hw}, and Seiberg duality
in four-dimensional $N=1$ supersymmetric YM theory with matter
\cite{EGK}, but there are many more \cite{GK}. The former follows
rather trivially from S-duality in Type IIB string theory, by considering
a configuration of parallel D3-branes suspended between two parallel 
NS5-branes. The low-energy effective field theory on the world-volume
of the D3-branes is then given by three-dimensional $N=4$ SYM theory.

It is well known that three-dimensional gauge theories can include,
in addition to the usual YM term, a Chern-Simons (CS) term 
in the action. The latter gives a gauge-invariant mass 
to the vector field proportional to the CS coefficient $k$.
In the (non-compact) abelian case $k$ is arbitrary, but
for non-abelian gauge groups invariance of the classical theory 
under large gauge transformations requires $k$ to be an integer \cite{jackiw}.
Like the pure YM theory, the combined YM-CS theory 
can be extended to include various amounts of supersymmetry, 
by including massive spinor and scalar fields transforming in the adjoint
representation of the gauge group. 
Unlike the pure YM case however, for which one can have as many as sixteen
supersymmetries, {\em i.e.} $N=8$, the largest amount of supersymmetry 
allowed in the YM-CS theory is $N=3$, {\em i.e.} six supersymmetries
\cite{lee}. Type IIB brane configurations for a class of supersymmetric
YM-CS theories which includes the $N=3$ theory have been constructed
in \cite{ohta1}. These configurations have been used to study various
non-perturbative aspects of SYM-CS theories, such as the structure of the
moduli space \cite{ohta2}, the spectrum of solitons \cite{korea},
and S-duality \cite{ohta3}.
An interesting question is whether, and under what conditions, 
supersymmetry is dynamically broken in such theories.
In \cite{ohta1} Kitao, Ohta and Ohta noted that 
for small enough $k$ some of the above configurations are not 
supersymmetric. 

Recently Witten has argued that for $N=1$ supersymmetric YM-CS theory
with gauge group $SU(n)$ and CS coefficient $k$ supersymmetry is
spontaneously broken when $|k|<n/2$ \cite{wCS}. 
This argument is based on a computation of the supersymmetric index 
$\Tr(-1)^F$, which vanishes in the above regime. 
For large $|k|$ one can integrate out the fermions
to obtain a local effective field theory, and at low energy 
one can ignore the YM term.
The resulting theory is pure CS with a shifted coefficient, 
given by\footnote{Note that for $n>1$ and odd this implies that $k$
must be quantized in half-odd-integer rather than integer units.
This is known as the parity anomaly \cite{walv,redlich}.}
\be
 k' = k - {n\over 2}\mbox{sgn}(m_f) \,,
\ee
where $m_f$ is the coefficient of the fermion mass term in the action.
Pure CS theory with gauge group $G$ and coefficient $k'$ is in turn related 
to the WZW model of $G$ at level $k'$. The correspondence implies in
particular that the number of zero-energy states in the CS theory is 
equal to the number of conformal blocks of the WZW model, and that these
states are all bosonic (or all fermionic). 
The index can therefore be computed by counting the number of 
conformal blocks, which for the above case gives
\be
I(k)={1\over (n-1)!}\prod_{j=-n/2+1}^{n/2-1}(k-j)\,.
\label{index}
\ee
In the regime where this computation is reliable, 
namely large $|k|$, the index is non-vanishing and 
supersymmetry is unbroken.
Note however that for $|k|<n/2$ the above expression vanishes, 
which suggests that supersymmetry may be spontaneously broken in this regime. 
Of-course, the approximation used to derive (\ref{index}) is not reliable
in this regime, and one must consider the full
microscopic theory.

The 
microscopic
problem can be studied by compactifying on a torus ${\bf T}^2$,
whose volume is sufficiently small, so that one can ignore the massive 
Kaluza-Klein modes \cite{wCS}. The index can then be computed by quantizing
the space of classical zero energy configurations.\footnote{
This includes all states with energies of order $g^2k$ or less.}
For the gauge field this is
the space of flat connections on ${\bf T}^2$, which for $SU(n)$ 
is given by a copy of ${\bf CP}^{n-1}$. 
There are also fermionic zero-modes with values in some bundle over 
${\bf CP}^{n-1}$.  
Quantization gives rise to a
Hilbert space of supersymmetric states, which is given by the
cohomology of the bundle over ${\bf CP}^{n-1}$. The result is an index given by
\be
 I(k) = \sum_{i=0}^n (-1)^i\, \mbox{dim}\,
   H^i({\bf CP}^{n-1},{\cal L}^{k-n/2}) \,,
\ee
where ${\cal L}$ is the basic line bundle over ${\bf CP}^{n-1}$.
Remarkably, this reduces to precisely the low-energy expression
for the index (\ref{index}).
More importantly, since all the cohomology groups are trivial for 
$|k|<n/2$, there are no zero energy states at all in this regime. 
Thus supersymmetry is spontaneously
broken for $|k|<n/2$, at least for a sufficiently small torus.
This suggests that the same is true in the infinite volume limit,
however a direct argument for this is not yet available.

A natural direction in which one might try to extend Witten's 
result is to consider more supersymmetry, {\em i.e.} $N=2$ and $N=3$.
In either case one introduces scalar fields, as well as additional
fermions, into the theory. 
Since the mass of these scalars is 
$g^2 k$,
they contribute to the space of classical zero-energy 
configurations, and thereby make the microscopic computation of the index 
much more involved than above. 
At large $|k|$ one can still use the low energy effective theory
obtained by integrating out the fermions 
(and scalars),
and the resulting index is
given by
\be
I(k)={1\over (n-1)!}\prod_{j=1}^{n-1}(k-j)\,,
\label{index2}
\ee
for both the $N=2$ and $N=3$ theories.
This differs from the $N=1$ result (\ref{index}) due the presence
of additional fermions.
One might however guess that, as in the $N=1$ theory, the large $|k|$
result would hold for all $|k|$, and therefore that
supersymmetry would be spontaneously broken in the $N=2$ and
$N=3$ theories (on ${\bf T}^2$) for $|k| < n$. 

\medskip

It is the purpose of this paper to establish that supersymmetry is 
indeed broken in the above regime for the $N=3$ theory, and for certain 
$N=2$ and $N=1$ deformations thereof. Our approach will be to make use of
recently proposed Type IIB brane configurations for these theories,
and show that these configurations break supersymmetry in the above regime.
The relevant configurations were constructed in \cite{ohta1}, and consist
of D3-branes suspended between an NS5-brane and a $(p,q)$ 5-brane.
In section 2 we will review the argument presented in \cite{ohta1}, 
which shows that the
$(p,q)$ 5-brane gives rise to a CS term with coefficient $k=p/q$ in
the low energy field theory of the suspended D3-branes. 
Note that for multiple D3-branes and $|q|>1$ this result is in conflict
with the requirement that $k$ be quantized in the non-abelian case.
We will partially resolve this conflict by providing an alternative
derivation of the above result, which is valid only for $|q|=1$. 
Specifically, the action contains additional terms when $|q|>1$.

The above configurations can be thought of as deformations of the 
configuration studied in \cite{hw}, which consisted of D3-branes 
suspended between two parallel NS5-branes. The latter corresponds 
to $D=3$ $N=4$ SYM theory.
By rotating the second 5-brane in both the spatial directions and
the eleventh ``M'' direction we produce a rotated $(p,q)$ 5-brane,
and break some of the supersymmetries. 
Depending on the relative orientation of the two 5-branes,
one obtains $N=3$ SYM-CS theory, or $N=2$ and $N=1$ deformations of this
theory, which include one and three matter multiplets, respectively. 
Note that the pure $N=1$  theory cannot be obtained by this 
construction.

We review the supersymmetric brane configurations and the corresponding
gauge theories in section 3. We then show that when the number of suspended
D3-branes $n$ exceeds a certain bound supersymmetry is in fact broken.
In most cases, and in particular for the $N=3$ theory, this is a 
consequence of the so-called ``s-rule'' \cite{hw}, which we review
in section 3 as well. In these cases we find that supersymmetry is
broken when $|k|<n$, precisely as predicted by the low energy
computation of the index.

In one special case, namely the $N=1$ deformation
with $k=0$, supersymmetry is  broken when $n>1$.
This can be understood as arising from an instanton in the
brane configuration, which gives rise to a repulsive potential
between the D3-branes. The $N=1$ theory can be thought of as
a mass deformation of an $N=2$ SYM theory, in which the instanton
gives rise to a superpotential. The mass deformation then results
in a potential with a non-vanishing minimum. 
This implies that three-dimensional $N=1$ SYM theory with 
gauge group $SU(n)$ and three flavors breaks supersymmetry when $n>1$.
The same can clearly not happen in the $k=0$ theories with higher
supersymmetry ($N=2,4$), as these correspond to dimensional reductions
of four-dimensional supersymmetric ($N=1,2$) YM theories, in which
supersymmetry is not broken. Since the $N=1$ theory cannot be obtained
by dimensional reduction from a supersymmetric theory in four dimensions,
there is no contradiction.

\section{CS terms and (p,q) 5-branes}

We begin with the Type IIB brane configuration studied in \cite{hw},
namely two parallel NS5-branes stretched along $(x^1,x^2,x^3,x^4,x^5)$
and separated along $x^6$, as well as $n$ D3-branes stretched between
them along $(x^1,x^2,x^6)$. In the low energy limit, where one can
ignore modes along the compact direction $x^6$, this gives a 
three-dimensional $N=4$ SYM theory with gauge group $U(n)$.
The $N=4$ vector multiplet consists of the vector field, four Majorana
fermions, and three real scalar fields, all transforming in the adjoint
representation of $U(n)$.  
The three scalar fields correspond to the positions of the D3-branes
in $x^3,x^4$, and $x^5$. 
Theories with less supersymmetry can be obtained by rotating one
of the NS5-branes in the $(x^3,x^7)$, $(x^4,x^8)$, or $(x^5,x^9)$ plane.
In the field theory on the D3-branes this translates into mass terms
for the three scalar fields proportional to the tangents of the three 
rotation angles.

Since in three dimensions the vector field is dual to
a scalar field, it would be natural to associate a similar mass
term to the vector field.
This is realized most naturally by lifting the above configurations
to M theory. 
Recall that Type IIB string theory corresponds to the compactification
of M theory on a vanishingly small torus.
Let us assume that the two 
compact coordinates of the torus are $\widehat{x}^2,x^{10}$.
 From the point of view of M theory, the original configuration 
then corresponds to two parallel M5-branes along
$(x^1,\widehat{x}^2,x^3,x^4,x^5)$, and $n$ M2-branes stretched
between them along $(x^1,x^6)$.
Rotating one of the M5-branes in the $(x^3,x^7)$, $(x^4,x^8)$,
and $(x^5,x^9)$ planes then reduces to the above rotated NS5-brane
configurations.
On the other hand, the M5-brane can also be rotated in the 
$(\widehat{x}^2,x^{10})$ plane. This will change the NS5-brane
into a $(p,q)$ 5-brane, with $p/q$ given by the tangent of the rotation angle.
Since, from the M theory point of view, the four angles are on an equal
footing, it is natural to guess that the angle of 
rotation in the internal torus will give mass to the fourth boson, 
namely the scalar field dual to the vector field of the $N=4$ vector
multiplet. 
This mass will again be proportional to the tangent of the angle, 
namely $p/q$. 
The only way to do this without breaking the gauge symmetry in the 
low energy field theory is to add a CS term 
with coefficient $k=p/q$ to the Lagrangian.
It is the purpose of this section to derive this result.

\subsection{Original derivation}

The content of the three-dimensional effective field theory on a D3-brane
suspended between 5-branes depends crucially on the boundary
conditions.
For an NS5-brane, the appropriate boundary condition is
\be
 {\bf NS5}: \quad 
  F_{\mu 6} = \pa_\mu A_6 - \pa_6 A_\mu = 0 \qquad (\mu=0,1,2)\,,
\label{ns5}
\nonumber
\ee
whereas for a D5-brane it is
\be
 {\bf D5}: \quad 
  F_{\mu \nu} = \pa_\mu A_\nu - \pa_\nu A_\mu = 0 \qquad (\mu,\nu=0,1,2)\,.
\label{d5}
\nonumber
\ee
These conditions remove some of the degrees of freedom of the 
four-dimensional gauge field.
In particular, the former removes the scalar component in the dimensional
reduction on $x^6$ (and therefore the $D=3, N=4$ hypermultiplet), 
and the latter removes the vector component (and therefore the $D=3, N=4$
vector multiplet).
The boundary condition corresponding to a $(p,q)$ 5-brane is obtained
by applying the appropriate $SL(2,\Z)$ transformation, which gives
\be 
  {\bf (p,q)\, 5}: \quad
  \pa_\mu A_6 - \pa_6 A_\mu - a \frac{p}{q} \e_{\mu\nu\lambda} 
  \pa^\nu A^\lambda = 0 \,,
\label{pqboundary}
\ee
where $a$ is an arbitrary constant at this stage.
This boundary condition introduces an unwanted surface
term to the variation of the action with respect to $A_\mu$.
To cancel this term one needs to add a boundary term. It was
argued in \cite{ohta1} that this boundary term in the four-dimensional theory
on the interval reduces 
to the CS action in the three-dimensional effective field theory
\be
\frac{a}{{g_4}^2} \frac{p}{q} \int d^3 x\, \e_{\mu\nu\lambda}
 A^\mu \pa^\nu A^\lambda \,.
\label{cs2}
\nonumber
\ee
The constant $a$ was then determined to be $g_4^2/4\pi$ 
by comparing with a more heuristic
derivation using the axion background of the $(p,q)$ 5-brane.
The coefficient of the CS term is therefore given by $k=p/q$.

A naive generalization to $n$ D3-branes would suggest a non-abelian
CS term with coefficient $k=p/q$. However, while fractional CS
coefficients are perfectly sensible in the abelian case
(and in fact give interesting predictions about a
$k \rightarrow 1/k$ mirror symmetry,
in accord with field theory expectations \cite{KS,ohta2,korea}),
it is well known that invariance under large gauge transformations
(in ${\bf R}^3$) requires $k$ to be an integer in the non-abelian case
\cite{jackiw}.
For $|q|=1$ the coefficient is an integer, and the CS term is
gauge invariant. 
It is not clear however where the above derivation fails if $|q|>1$.

In order to gain insight into this issue, we shall offer an alternative
derivation of the CS term, which reproduces the above result for $|q|=1$,
and clearly demonstrates the complications associated with $|q|>1$.

\subsection{Alternative derivation}

Our strategy will be to identify the $(p,q)$ 5-brane as a particular
mass deformation of 
an abelian 
three-dimensional gauge theory, and then show
that this mass deformation induces the desired CS term. This will follow 
from the
basic fact that integration of massive fermions in three dimensions yields an
effective CS term which arises in a one loop computation \cite{redlich, lee}.
A CS term is generated for each pair of $U(1)$ groups present
in the problem, whether they be local or global.
In either case one will get an interesting effect.
The contribution to a given
CS term is proportional to the sign of the mass of the fermion and to the
charges of the fermion under the two $U(1)$ groups in question.
For a fermion of mass $m_f$ and charges $q_i$ the contributions 
to the CS terms are therefore of the form
\be
{1\over 2}q_iq_j\mbox{sgn}(m_f).
\ee

\begin{figure}
\refstepcounter{figure}
\label{real}
\begin{center}
\makebox[8cm]{
     \epsfxsize=8cm
     \epsfysize=5cm
     \epsfbox{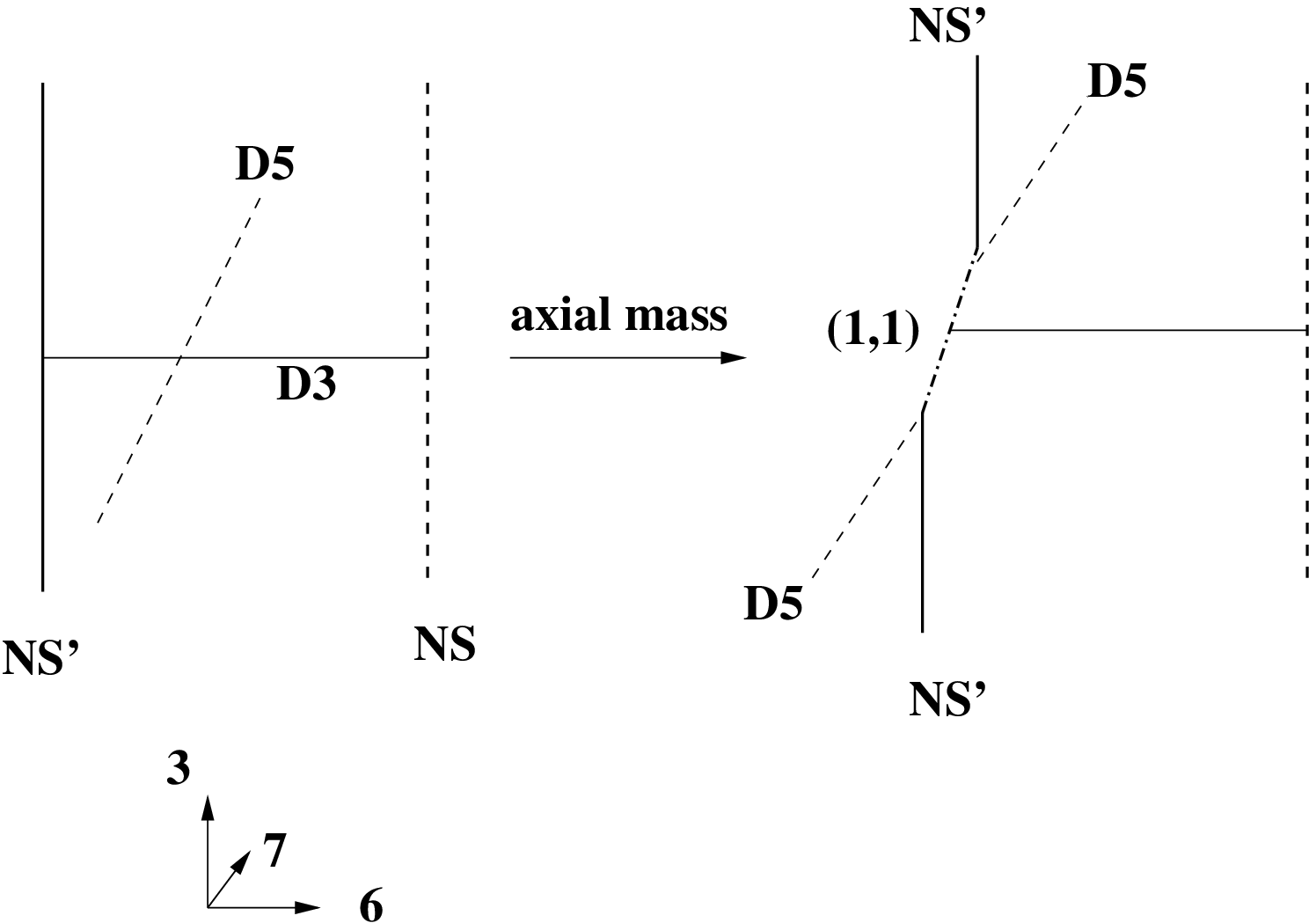}
}
\end{center}
\center{{\bf Fig.\thefigure}{\bf.}
Turning on an axial mass for the electron.}
\end{figure}

The theory we shall consider is three-dimensional $N=2$ supersymmetric 
$U(1)$ Maxwell theory with one flavor of matter.
This theory corresponds to the dimensional reduction of four-dimensional
$N=1$ Maxwell theory with one flavor. 
The massless vector multiplet consists of a vector field, two Majorana 
fermions and a scalar field. The matter consists of two oppositely
charged chiral multiplets
$Q,\tilde{Q}$ (``electron'' and ``positron''),
each containing a two-component Majorana fermion 
and a complex scalar.
This theory can be realized on a single D3-brane
suspended between an NS5-brane along $(x^1,x^2,x^3,x^4,x^5)$
and another (NS5$'$) along $(x^1,x^2,x^3,x^8,x^9)$ \cite{EGK}.
The matter multiplet is provided by a D5-brane along $(x^1,x^2,x^7,x^8,x^9)$
(See Fig.\ref{real}).

There are two possible mass terms for matter in three dimensions 
\cite{5authors}.
The first is inherited from the four-dimensional $N=1$ theory,
and corresponds to the standard complex mass parameter obtained from
the superpotential,
\be
 \int d^2\theta m_c \tilde{Q}{Q} \,.
\ee
In the brane picture this corresponds to the
position of the D5-brane in the $(x^4,x^5)$ plane.
The second is a real mass term of the form
\be
 \int d^4 \theta Q^{\dagger} e^{{m} \theta \bar{\theta}} Q \,,\quad
 \int d^4 \theta \tilde{Q}^{\dagger} e^{\tilde{m}\theta\bar{\theta}}
   \tilde{Q}\,,
\label{realmass}
\ee
which can be chosen independently for the two chiral multiplets. 
We define the axial and vector masses respectively as
\be
 m_A = {1\over 2} (m+\tilde{m})\,,\quad
 m_V = {1\over 2} (m-\tilde{m})\,.
\ee
Since the minimal coupling of the charged matter fields to the
vector multiplet takes the standard form
\be
  \int d^4 \theta Q^{\dagger} e^{V} Q \,,\quad
  \int d^4 \theta \tilde{Q}^{\dagger} e^{-V} \tilde{Q}\,,
\label{minimal}
\ee
we see that the vacuum expectation value (VEV) of the scalar in the $N=2$ 
vector multiplet gives a real mass of equal magnitude and opposite
sign to the two chiral multiplets.  
By shifting this scalar field we can therefore set $m_V=0$.
The total mass terms for the electron and positron would then be
given by $(\Phi + m_A)|Q|^2$ and $(-\Phi + m_A)|Q|^2$, respectively.

In the brane picture the VEV of the scalar $\Phi$ corresponds to the 
$x^3$ position of the D3-brane, whereas the real mass parameter
$m_V$ corresponds to the $x^3$ position of the D5-brane.
At this point the attentive reader will wonder how the axial
mass $m_A$ comes about.
In the brane picture we must somehow break the D5-brane into two independent
pieces, whose positions in $x^3$ will correspond to the mass parameters
of the two fermions. Following \cite{brodie}, this is done by moving the 
D5-brane along the compact direction $x^6$ until it coincides with the 
NS5$'$-brane, at which point it can break along $x^3$ into 
two D5-branes which end
on the NS5$'$-brane. 
The two fermions (chiral multiplets) then correspond to strings between 
the D3-brane and the two pieces of D5-brane.
In order to preserve supersymmetry, the 5-brane system has to form
a $(p,q)$ web in the $(x^3,x^7)$ plane consisting of the two pieces of 
D5-brane, two pieces of NS5$'$-brane, and an intermediate $(1,\pm 1)$ 
5-brane,
where the sign depends on which way the D5-brane breaks \cite{ofer}.
Turning on an axial mass corresponds to moving the two D5-brane pieces
an equal amount, but in opposite directions, 
along $x^3$ (see Fig.\ref{real}),
and the sign of the mass is correlated with the sign in the intermediate
5-brane. 

As there are two $U(1)$ groups in the problem, namely the gauge group
and the global axial symmetry, integrating out the fermions
produces two relevant CS terms. The first is the usual CS term for
the gauge field with coefficient
\be
{1\over2}\mbox{sgn}(m_A+\Phi)+{1\over2}\mbox{sgn}(m_A-\Phi)\,,
\label{csgen}
\ee
and the second is a mutual CS term for the two $U(1)$'s
with coefficient
\be
{1\over2}\mbox{sgn}(m_A+\Phi)-{1\over2}\mbox{sgn}(m_A-\Phi)\,.
\label{figen}
\ee
The latter corresponds to an FI term in the action.

As we vary $\Phi$ the D3-brane moves along $x^3$.
For $\Phi > m_A$ and $\Phi < -m_A$  the D3-brane ends on the ordinary 
NS5$'$-brane, which is
displaced by $\pm m_A$ along $x^7$. The mass terms have opposite signs,
so the CS coefficient (\ref{csgen}) vanishes. On the other hand there
is a non-trivial FI term (\ref{figen}), which is consistent with the
displacement.
For $-m_A<\Phi<m_A$ the D3-brane ends on the $(1,\pm 1)$ 5-brane, 
and the mass terms have the same sign. The net CS coefficient is 
therefore given by $k = \pm 1$. This establishes the result we were
after for the case $p=\pm q=1$.

\subsection{Generalizations}

The above argument is easily generalized for $U(n)$ by including $n$
D3-branes instead of just one.
One can also generalize to an arbitrary integer CS coefficient $k=p$
by including $p$ D5-branes (and thus $N_f=p$ flavors). Turning on
an axial mass for all the flavors will correspond in the brane
picture to a web of $p$ D5-branes, 1 NS5$'$-brane, and an intermediate 
$(p,1)$ 5-brane.
The contributions of the $p$ flavors then simply add up to a CS coefficient 
$k=p$, as anticipated.

In order to further generalize to fractional CS coefficients $p/q$
($|q|>1$), which are supposed to arise from a $(p,q)$ 5-brane,
we need to start with multiple ($|q|$) coinciding NS5$'$-branes.
It turns out, however, that the corresponding effective field theories on
branes suspended between the NS5-brane and NS5$'$-branes 
contain additional matter \cite{EGK}. 
For example in four dimensions, {\it i.e.} on D4-branes suspended
between an NS5-brane and several NS5$'$-branes in Type IIA,
the IR limit of the theory is described by SYM with $N_f$ fundamental flavors
(from D6-branes)
coupled to an additional chiral multiplet $X$ transforming in the adjoint
representation, with a superpotential of the form
\be
 W=X^{|q|+1} + Q X \tilde{Q} \,.
\ee
This is a rather complicated theory. In particular, its
Seiberg-dual is presumably some kind of tensionless string theory
\cite{kutasov}. 

In three dimensions we also expect additional massless adjoint fields,
although the precise description of the IR physics is not known.
The presence of these fields should give rise to new interesting effects, 
such as an enhanced global $SU(|q|)$ symmetry from the coinciding 
NS5$'$-branes.
Upon turning on the real mass terms for all the (fundamental) flavors
as before, the vector multiplet would again become massive. This
time, however, the IR dynamics would be governed by the remaining massless
adjoint multiplet and its self-interactions, rather than by the CS term.
We therefore avoid getting pure CS dynamics with non-integer coefficient.

Of course in the abelian case the CS coefficient can be fractional 
(and in fact real).
It would be interesting to understand this in the above context,
especially since the brane configuration for a non-integer CS coefficient
\cite{ohta1}, and the corresponding spectrum of vortices \cite{korea,ohta2},
predicts the $k\rightarrow 1/k$ duality of \cite{KS}.
In the abelian case the adjoint multiplet $X$ is neutral, and
couples only to the charged fields. 
One might hope that in the IR
it decouples altogether, so that the analysis of \cite{ohta1} makes
sense, and we can 
associate 
a Maxwell-Chern-Simons theory with
this system, where $k=p/q$. 

\medskip

As a final generalization, consider a second type of flavor D5-brane
(D5$'$), which is stretched along $(x^1,x^2,x^4,x^5,x^7)$, in addition
to the original type along $(x^1,x^2,x^7,x^8,x^9)$.
This leads generically to quartic superpotentials \cite{ofer},
whose precise form depends on the ordering of the two types of D5-branes.
If all the D5-branes are to the left of all the D5$'$-branes
(so that the former can be moved towards the NS5$'$-brane without crossing
the latter, and the latter can be similarly moved towards the NS5-brane)
there is no superpotential.
Turning on axial masses for all the flavors then corresponds to
5-brane webs in the $(x^3,x^7)$ plane involving a $(p_1,\pm 1)$ 5-brane 
at one end and a 
$(p_2,\pm 1)$ 5-brane at the other end. 
 From the field theory point of view all $p_1+p_2$
flavors are the same, so the result is identical to the one obtained
from $p_1+p_2$ D5-branes, namely a CS coefficient $k=\pm p_1 \pm p_2$.
The signs are somewhat ambiguous at this stage. To fix this ambiguity
we use the fact that the configuration of D3-branes between two
parallel $(1,1)$ 5-branes preserves $N=4$ supersymmetry, which forbids
a CS term. So for a $(p_1,1)$ 5-brane and a $(p_2,1)$ 5-brane
the CS coefficient is $k=\pm(p_1-p_2)$.
For multiple NS5-branes at both ends we therefore get
\be
 k = \pm {p_1q_2 - p_2q_1\over q_1q_2} \,.
\ee
As before, the low energy theory will contain additional massless
fields if $q_1>1$ or $q_2>1$, which one would hope decouple in the abelian 
case.

Of course now that one has altered the boundary condition at the other
end, the original hypermultiplet is no longer removed, but in fact
obtains a finite mass, which is complementary to the mass of the vector
multiplet, namely
\be
 m_h = \left|{p_1q_2 - p_2q_1\over p_1p_2}\right| \,.
\ee
This gives the required massless hypermultiplet if both 5-branes
are D5-branes.

\section{SUSY brane configurations and SUSY breaking}

\subsection{Brane configurations and related field theories}

As we are interested in theories containing only the original 
three-dimensional $N=4$ vector multiplet, we shall only consider 
configurations of D3-branes suspended between an NS5-brane and
a $(p,q)$ 5-brane.
All possible supersymmetric configurations of 
this type have been classified in \cite{ohta1}, and are summarized in 
table~\ref{t1}. The rotation angles, from the 
M theory 
point of view,
in the $(\widehat{x}^2,x^{10})$, $(x^3,x^7)$, $(x^4,x^8)$, and
$(x^5,x^9)$ planes are taken to be $\theta$, $\psi$, $\varphi$, and $\rho$,
respectively. In particular, $\theta$ determines $(p,q)$ as
\be
 \tan{\theta} = {g_sp\over q} \,,
\ee
and the other angles determine the masses of the three scalar fields
corresponding to the position of the D3-brane in $(x^3,x^4,x^5)$.
If any of the latter angles vanish, as in configurations 1, 2(i,ii), and 3(ii),
the theory contains massless scalars, and the CS term is no longer the dominant
part of the IR dynamics.
Therefore we are only really interested in the cases
in which all scalar fields are massive, {\it i.e.} 3(i) and 4(i,ii,iii).

In particular, 4(iii) corresponds to the maximally supersymmetric $N=3$ 
SYM-CS theory with CS coefficient $k=p/q$. 
The associated R-symmetry is $SO(3)$, and
the {\em massive} vector multiplet consists of a spin 1 vector field 
$A_\mu$, three spin 1/2 Majorana fermions $\lambda_a$, 
one spin -1/2 fermion $\chi$, and three scalars $\phi_a$.\footnote{This is 
a {\em long} rather than {\em short} (BPS) multiplet. 
Since the mass terms come from a CS term rather than from spontaneous 
symmetry breaking, the central charge of the superalgebra vanishes.}
The YM and CS Lagrangians of
this theory are respectively given by \cite{lee}
\begin{eqnarray}
L_{YM} &=& {1\over 4g^2}\int d^3x\, \Tr\, \bigg\{-F^2 +
   (D_\mu\phi_a)^2 + (C_a)^2
  + i\overline{\lambda}_a\Gamma^\mu D_\mu\lambda_a
  + i\overline{\chi}\Gamma^\mu D_\mu\chi  \nonumber \\
  && \mbox{} + i\epsilon_{abc}\overline{\lambda}_a[\lambda_b,\phi_c]
   - 2i\overline{\lambda}_a[\chi,\phi_a]
   +{1\over 2}[\phi_a,\phi_b]^2 \bigg\}
\end{eqnarray}
and
\begin{eqnarray}
 L_{CS} &=& {k\over 4\pi}\int d^3x\,\Tr\,\bigg\{
   \epsilon^{\mu\nu\rho}(A_\mu \del_\nu A_\rho 
   - {2\over 3}i A_\mu A_\nu A_\rho)
   - \overline{\lambda}_a\lambda_a + \overline{\chi}\chi
   + 2\phi_aC_a \nonumber \\
  && \hfil \mbox{} + {i\over 3}\epsilon_{abc}\phi_a[\phi_b,\phi_c]
   \bigg\}\,,
\end{eqnarray}
where $C_a$ are auxiliary scalar fields. 
All the fields have the same mass $g^2k$, but
the mass term for $\chi$ enters with a sign opposite to that of $\lambda_a$. 
The YM Lagrangian by itself actually 
has $N=4$ supersymmetry, with an $SO(4)$ R-symmetry, and the vector 
multiplet is massless.\footnote{The massless three-dimensional $N=4$ vector 
multiplet is given in terms of $SO(4)\sim SU(2)\times SU(2)$
representations by
$({\bf 1},{\bf 1})\oplus ({\bf 3},{\bf 1})\oplus ({\bf 2},{\bf 2})$,
corresponding to the vector, three scalars, and four fermions, respectively.
This has the same content as the massive $N=3$ vector multiplet.}
The different mass terms for the fermions in the CS Lagrangian
break 
the supersymmetry to $N=3$, and the R-symmetry to $SO(3)$.
The total scalar potential is given by
\be
 V = {1\over 4g^2} Tr \left(\frac{1}{3\pi} g^2k \phi_a - \frac{1}{2} 
   \epsilon_{abc} [\phi_b,\phi_c]\right)^2
    + \frac{8}{9\pi^2} g^2k^2 \phi_a^2 \,,
\ee
so all the flat directions of the $N=4$ theory are lifted.

The configurations 4(ii) and 4(i) are deformations of 4(iii), whereby
one changes the spatial angles $\psi, \phi$, and $\rho$  
relative to $\theta$. These correspond
therefore to $N=2$ and $N=1$ deformations of the $N=3$ theory, in
which one adds additional mass terms to the fields $\phi_{1,2}, \lambda_3$
and $\chi$ in the $N=2$ case, and to $\phi_3$ and $\lambda_2$ as well
in the $N=1$ case.
Configuration 3(i) corresponds to the special case $k=0$,
{\it i.e.} $\theta = 0$, of the $N=1$ deformation.
In this theory the gauge field (and its fermion superpartner $\lambda_1$) 
is massless, and the three matter multiplets containing the fermions 
$\chi, \lambda_2$, and $\lambda_3$, as well as all the scalars, are massive.

\begin{table}[htb]
\begin{tabular}{|c|c|c|c|c|l|}
\hline
 configuration & angles & condition & SUSY & $D=3$ & second 5-brane\\
\hline
\hline
 &&&&& \\[-13pt]
1 & $\theta$ & $\theta=0$ & 1/4 & $N{=}4$ & NS5 $(12345)$ \\[1pt]
\hline
 &&&&& \\[-13pt]
2(i) & $\varphi, \rho$ & $\rho=\varphi$ & 1/8 &  $N{=}2$
 & NS5 $\Big(123\rot{4}{8}{\varphi}\rot{5}{9}{\varphi}\Big)$ \\[5pt]
\hline
 &&&&& \\[-13pt]
2(ii) & $\theta, \rho$ & $\rho=\theta$ & 1/8 & $N{=}2$
 & $(p,q)5 \Big(1234\rot{5}{9}{\theta}\Big)$ \\[5pt]
\hline
 &&&&& \\[-13pt]
{\bf 3(i)} & $\psi, \varphi, \rho$ & $\rho=\psi+\varphi$ 
 & 1/16 & $N{=}1$ & 
 NS5 $\Big(12\rot{3}{7}{\psi}\rot{4}{8}{\varphi}\rot{5}{9}{\psi+\varphi}
   \Big)$ \\[5pt]
\hline
 &&&&& \\[-13pt]
3(ii) &  $\theta, \varphi, \rho$ & $\rho=\theta+\varphi$
 & 1/16 & $N{=}1$ & 
 $(p,q)5 \Big(123\rot{4}{8}{\varphi}\rot{5}{9}{\theta+\varphi}\Big)$ 
  \\[5pt]
\hline
 &&&&& \\[-13pt]
{\bf 4(i)} & 
 & $\rho=\theta+\psi+\varphi$ & 1/16 & $N{=}1$ & $(p,q)5
\Big(12\rot{3}{7}{\psi}\rot{4}{8}{\varphi}
\rot{5}{9}{\rho}\Big)$ \\[5pt]
\cline{1-1}\cline{3-6}
 &&&&& \\[-13pt]
{\bf 4(ii)} & $\theta, \psi, \varphi, \rho$
 & $\varphi=-\psi, \rho=\theta$ & 1/8 & $N{=}2$
 & $(p,q)5\Big(12\rot{3}{7}{\psi}\rot{4}{8}{-\psi}\rot{5}{9}{\theta}\Big)$
 \\[5pt]
\cline{1-1}\cline{3-6}
 &&&&& \\[-13pt]
{\bf 4(iii)} & & $\theta=\rho=\varphi=-\psi$ & 3/16 & $N{=}3$
 & $(p,q)5\Big(12\rot{3}{7}{-\theta}\rot{4}{8}{\theta}
 \rot{5}{9}{\theta}\Big)$ \\[5pt]
\hline
\end{tabular}
\caption{\small
Brane configurations and supersymmetric gauge theories in three dimensions.}
\label{t1}
\end{table}

As we have seen in the previous section,
the derivation of the CS coefficient was only reliable 
in the case $|q|=1$. We shall therefore restrict ourselves to this case,
and thus assume that $k=p$.

\subsection{SUSY breaking via the s-rule}

The amount of supersymmetry listed for each of the above configurations
was determined by intersecting the supersymmetries preserved by each
individual brane. This is a reliable procedure for counting 
supersymmetries when the branes are all infinite in extent.
When some of the branes are suspended between others, on the other hand,
supersymmetry can sometimes be broken if the number of suspended
branes exceeds a certain value, even though the analogous
configuration of infinite branes is supersymmetric.
The first example of this phenomenon was given in \cite{hw},
and involved a configuration of D3-branes suspended between 
an NS5-brane and a D5-brane, which had precisely one common transverse
direction. This configuration appears to preserve eight (1/4) supersymmetries.
However, it was argued that in order to reproduce
known field theory results one had to assume that the above configuration
breaks supersymmetry if more than one D3-brane is present.
This apparently empirical rule was termed the ``s-rule''.

The s-rule was subsequently derived in string theory for brane configurations
which are related to the above by a series of dualities, and which 
consist of a D$p$-brane and a D$(8-p)$-brane which are mutually
transverse, together with fundamental strings suspended between them
\cite{bg}. Each open string between the two D-branes has
a non-degenerate ground state consisting of a single fermion.
As the D-branes are completely transverse to each other, these fermions
are localized at a point, and are therefore subject to the Pauli exclusion
principle \cite{bg}. Hence, only one string can be in its supersymmetric ground
state. The others must be in non-supersymmetric excited states.
Therefore a configuration with more than one string breaks supersymmetry.

Another way to see this is by the string creation phenomenon
\cite{bf}, which occurs in the above configurations when the D-branes 
cross each other \cite{bdg,dfk,bgl1}. 
Assume there are $n$ suspended strings initially.
Denoting the 16-component supercharges of Type II string theory
by $Q_L$ and $Q_R$, the (apparent) unbroken supersymmetry of the above 
configuration is generated by $\epsilon_L Q_L + \epsilon_R Q_R$,
where the spinors $\epsilon_L$ and $\epsilon_R$ are solutions of
the following equations
\be
\begin{array}{ll}
 \mbox{Dp-brane} & \epsilon_L = \Gamma^1 \cdots \Gamma^p \epsilon_R \\
 \mbox{D(8-p)-brane} & \epsilon_L = \Gamma^{p+1}\cdots \Gamma^8\epsilon_R \\
 \mbox{string} & \epsilon_L = \Gamma^0\Gamma^1\epsilon_L \;, 
                 \epsilon_R = \pm \Gamma^0\Gamma^1\epsilon_R\;,
\end{array}
\ee
and the sign in the last equation depends on the orientation of
the suspended string. There exists a non-trivial solution only
for the upper sign, which means that one orientation preserves
(1/4) supersymmetry, while the other breaks it.
Let us assume therefore that the $n$ strings are oriented in the former 
way, so that the configuration appears to be supersymmetric. 
As the branes cross the orientation of the suspended strings is reversed,
and a single string is created with the original (supersymmetric) orientation.
The latter annihilates with one of the former, leaving $n-1$
strings with the non-supersymmetric orientation. 
Since the final configuration is not supersymmetric for $n>1$, 
neither is the initial one.
One can then use combinations of T-duality and S-duality to relate
the D$p$-D$(8-p)$-string system to various other systems, including
the NS5-D5-D3 system, and thereby establish the above supersymmetric bound 
on the number of suspended branes for these systems as well.

It is clear that the s-rule is intimately connected to the brane
creation phenomenon. This is in fact true also for the original
setting of D3-branes between NS5-branes and D5-branes of \cite{hw}.
This connection allows us in turn to generalize the s-rule
to brane configurations in which multiple suspended branes
are created, and which are not related by dualities to any of the 
above configurations.
For example, the system of a $(p,q)$ string and an $(r,s)$ 7-brane
exhibits the creation of $|ps-qr|$ $(r,s)$ strings, which implies
the s-rule $n\leq |ps-qr|$ \cite{bf}. 
Similarly, a configuration of a $(p,q)$ 5-brane and an $(r,s)$ 5-brane
which have precisely one common transverse direction exhibit the creation
of $|ps-qr|$ D3-branes \cite{ohta1}. 
This can be understood by decomposing each 5-brane
into its D5-brane and NS5-brane components, and applying the creation
phenomenon piecewise. The statement of the s-rule in this case is 
again $n\le |ps-qr|$.

For the configurations we are interested in, {\it i.e.} an NS5-brane
and a $(k,1)$ 5-brane oriented as in 4(i,ii,iii), supersymmetry is therefore
spontaneously broken when
\be
 n>|k| \,.
\ee
We conclude that supersymmetry is spontaneously broken in $N=3$
SYM-CS theory with gauge group $SU(n)$ when $n>|k|$, and similarly in its
$N=2$ and $N=1$ deformations.

\subsection{SUSY breaking via instantons}

Other than configurations 4(i,ii,iii), the only remaining configuration 
which leads to a field theory with no massless scalars is 3(i).
This configuration is a special case of 4(i), in which both 5-branes
are NS5-branes, {\it i.e.} $(0,1)$ 5-branes.
The corresponding gauge theory has $k=0$,
and a naive application of the preceding s-rule formula seems to imply that
supersymmetry is broken for $n>0$, and in particular for $n=1$.
However, the latter is recognized as the {\em free} $N=1$ $U(1)$ 
super-Maxwell theory with three massive neutral matter multiplets.
Clearly supersymmetry cannot be broken in this case.

To resolve the issue, recall that the s-rule is a consequence of brane 
creation, together with the fact that only one of the two possible 
orientations
of the suspended brane preserves supersymmetry. In the above configuration
there is no brane creation, which on the face of it seems to imply
the aforementioned bound. On the other hand, both orientations of the
D3-brane preserve supersymmetry in this case, 
so our derivation of the s-rule does not apply.

For $n>1$ the theory is interacting, and supersymmetry might be
broken, albeit via a different mechanism than the s-rule.
To see that this is indeed the case, consider the $N=1$ configuration 3(i)
as a deformation of the $N=2$ configuration 2(i). 
The latter corresponds in the IR limit to $D=3$, $N=2$ SYM with
a single massive $N=2$ matter multiplet. The massless scalar field in the
vector multiplet corresponds to the $x^3$ positions of the D3-branes.
This field becomes massive in the deformation from 2(i) to 3(i).

Consider two D3-branes suspended between the two NS5-branes in the $N=2$
configuration.
A Euclidean D-string suspended between the four branes
on the $(x^3,x^6)$ plane then gives rise to a superpotential
\cite{ofer}
\be
 W = e^{-\delta x^3 L/g_s\alpha'} \,,
\ee
where $\delta x^3$ is the separation between the D3-branes along 
$x^3$, and $L$ is the separation between the NS5-branes along $x^6$.
There is therefore a repulsive interaction between the two D3-branes,
leading to a runaway behavior in $\delta x^3$.

Now deform the configuration by rotating the second NS5-brane in the
$(x^3,x^7)$ plane, while maintaining $N=1$ supersymmetry. This
corresponds to adding a mass term to the scalar
field (and one of the fermions) in the $N=2$ vector multiplet.
In particular, this implies a mass term for $\delta x^3$,
and thus a total effective potential given by
\be
 V(\delta x^3) = m^2(\delta x^3)^2 + e^{-2\delta x^3 L/g_s\alpha'} \,.
\ee
Since this has a non-degenerate minimum of finite non-vanishing energy,
supersymmetry is broken. This also follows from the fact that in the 
minimum energy configuration the D3-branes are separated a finite distance
in $x^3$, and can therefore not be parallel.
We conclude that in $N=1$ $SU(n)$ SYM with three massive flavors supersymmetry
is broken when 
\be
 n>1 \,.
\ee

\section{Conclusions}

Brane configurations have once again proven themselves useful in
deriving non-perturbative properties of supersymmetric quantum field
theories. Here we have demonstrated dynamical supersymmetry breaking
in a class of three-dimensional gauge theories with a Chern-Simons term,
by studying their realization on configurations of D3-branes suspended
between an NS5-brane and a $(p,q)$ 5-brane (at least for $|q|=1$).
These theories include $N=3$ supersymmetric $SU(n)$ YM-CS theory, as well
as its $N=2$ and $N=1$ deformations, which contain in addition
one and three adjoint matter multiplets, respectively.
Supersymmetry is broken in the corresponding configurations, via the
``s-rule'', when the number of D3-branes $n$ is larger than the CS 
coefficient $k$. This is in line with earlier interpretations
of the s-rule as spontaneous supersymmetry breaking. 

The above supersymmetry breaking regime is consistent with 
the low energy computation of the supersymmetric index, which vanishes
in precisely this regime. 
However the low energy result is reliable only for large $|k|$,
and one should really compare with a microscopic calculation.
Such a calculation, as was done by Witten for
the pure $N=1$ theory, is not yet available for the $N=3$ theory.
The above agreement suggests however that, as in the pure $N=1$
theory, the low energy result should hold microscopically.

The case $k=0$, {\em i.e.} no CS term, should be excluded from the 
supersymmetry breaking
regimes of the $N=3$ (which is now really $N=4$) and $N=2$
theories, since these correspond to the dimensional reduction of
four-dimensional SYM theory with $N=2$ and $N=1$, respectively.
This does not rule out supersymmetry breaking in the $N=1$ theory
with three massive flavors and $k=0$. 
Indeed we learn from the corresponding brane configuration
that supersymmetry is broken by an instanton generated repulsive
potential between the D3-branes when $n>1$.

There are two interesting issues which deserve future attention.
The first concerns the failure of interpreting $p/q$ as the CS coupling when
$|q|>1$. A non-integer CS coefficient should be allowed in the abelian
case, and the corresponding brane configuration would be crucial in deriving
the $k\rightarrow 1/k$ duality of \cite{KS}.
The second issue is whether there exists a brane configuration for 
the pure $N=1$ SYM-CS theory, for which a microscopic calculation of
the supersymmetric index is available \cite{wCS}. 
Possible places to look for these configurations are brane boxes and
brane cubes.

\section*{Acknowledgments}

We would like to thank Mina Aganagic, Ofer Aharony, Anton Kapustin, 
and Shimon Yankielowicz for useful conversations.
A.H. and B.K. thank the organizers of the Jerusalem workshop and the organizers of the SUSY99 program at the ITP in Santa Barbara for hospitality during different stages of this work.
O.B. is supported in part by the DOE under grant no. DE-FG03-92-ER40701. 
A.H. and A.K. are supported in part by the DOE under grant no.
DE-FC02-94ER40818.
A.H. is supported in part by an A.P. Sloan Foundation Fellowship and by
a DOE OJI award.
This research was supported in part by the National Science
Foundation under Grant no. PHY94-07194.

\bibliography{cs}
\bibliographystyle{utphys}

\end{document}